\documentclass[aps]{revtex4}
\usepackage{graphicx}
\usepackage{epsfig}
\usepackage{amsmath,amssymb}

\def\fig#1{Fig.~\ref{#1}}

\begin{document}

\title{Hysteresis in the cell response to time-dependent substrate stiffness}

\author{Achim Besser}
\affiliation{University of Heidelberg, Bioquant, Im Neuenheimer Feld 267, 69120 Heidelberg, Germany}
\author{Ulrich S. Schwarz}
\affiliation{University of Heidelberg, Bioquant, Im Neuenheimer Feld 267, 69120 Heidelberg, Germany}
\affiliation{University of Heidelberg, Institute for Theoretical Physics, Philosophenweg 19, 69120 Heidelberg, Germany}

\begin{abstract}
{Mechanical cues like the rigidity of the substrate are main
  determinants for the decision making of adherent cells. Here we use
  a mechano-chemical model to predict the cellular response to varying
  substrate stiffness. The model equations combine the mechanics of
  contractile actin filament bundles with a model for the
  Rho-signaling pathway triggered by forces at cell-matrix contacts.
  A bifurcation analysis of cellular contractility as a function of
  substrate stiffness reveals a bistable response, thus defining a
  lower threshold of stiffness, below which cells are not able to
  build up contractile forces, and an upper threshold of stiffness,
  above which cells are always in a strongly contracted state. Using
  the full dynamical model, we predict that rate-dependent hysteresis
  will occur in the cellular traction forces when cells are exposed to
  substrates of time-dependent stiffness.}
\end{abstract}

\maketitle

\section{Introduction}

During the last decade, it has been shown that all adherent cell types
sense and respond to mechanical cues in their environment, including
substrate stiffness and prestress \cite{Janmey09}. For example,
fibroblast-like cells adhere stronger to stiffer substrates, including
decreased motility, increase in spreading area and higher contractile
forces. Most strikingly, the fate of stem cells can be controled by
substrate stiffness \cite{Discher09}. It is becoming increasingly
clear that mechanosensing by tissue cells is not based on the
functioning of one particular molecular entity, but depends on the
dynamical function of a mechanosensitive system whose main components
are actomyosin force generation and signaling from cell-matrix
adhesions \cite{Geiger09}. For example, all the described effects are
abolished by inhibiting myosin II activity by blebbistatin or the
Rho-pathway by C3-toxin. Here we perform a systems-level analysis
of rigidity sensing by tissue cells taking into account both the
mechanical and biochemical aspects.

\begin{figure}
\includegraphics[width=0.7\textwidth]{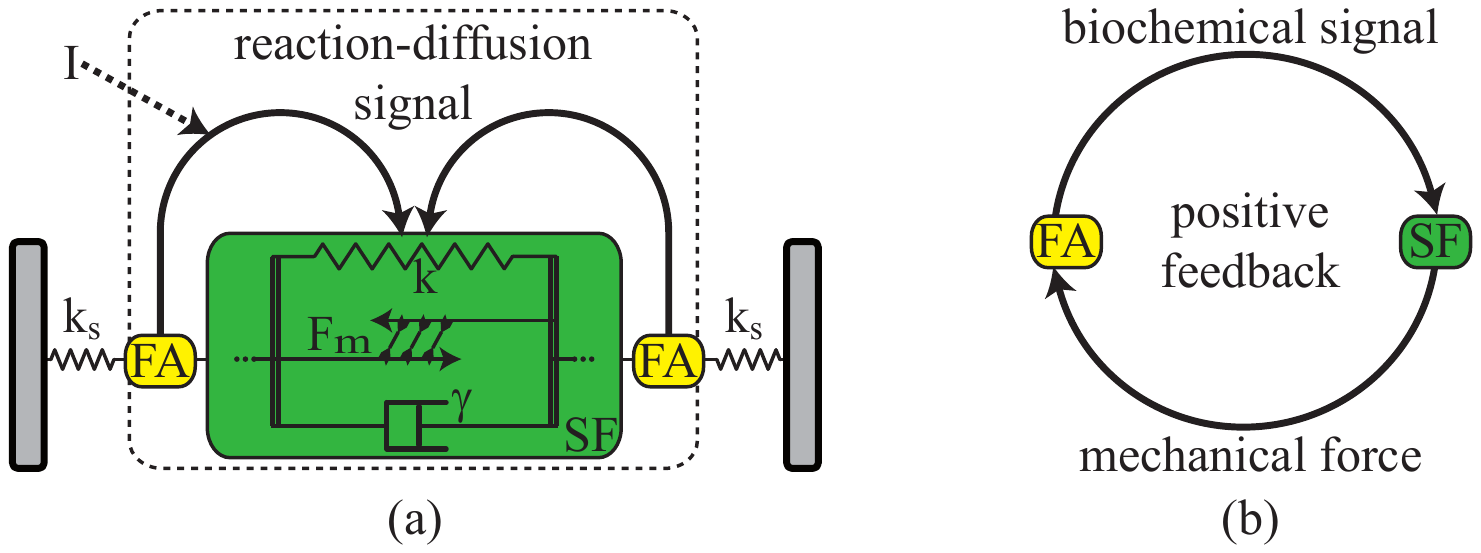}
\caption{The mechano-chemical model: (a) The stress fiber (SF) is
  modeled as a continuous one-dimensional Kelvin-Voigt material
  (spring stiffness $k$, viscosity $\gamma$) that may locally contract
  due to actomyosin forces ($F_m$).  Force-induced signaling at focal
  adhesions (FA) is described by a reaction-diffusion system
  (Rho-pathway). Substrate stiffness $k_s$ and biochemical stimulation
  $I$ (e.g. calyculin) are used as control parameters. (b) Higher
  forces on FAs increase Rho-signaling which in turn lead to higher
  myosin activation, thus closing a positive feedback
  loop.}
\label{fig_model}
\end{figure}

Our model is sketched in \fig{fig_model}(a). Cell mechanics is
represented by a viscoelastic model for a stress fiber (SF), which is
the most prominent feature of the actin cytoskeleton developed in cell
culture on flat substrates. The parallel arrangement of elastic and
viscous elements leads to a Kelvin-Voigt element, that is the SF
behaves like a solid on long time scales.  In addition to the passive
viscoelastic stress, we account for active actomyosin contractility.
The myosin forces are determined by a force velocity relation whose
properties are modulated by biochemical signals diffusing in the
cytoplasm. Therefore actomyosin contractility may vary spatially along
the fiber.  In the continuum limit of many elements in series, we
arrive at a continuous Kelvin-Voigt material governed by a partial
differential equation with mixed derivatives, the stress fiber equation
\cite{Besser07}.

The forces generated in the SF are transmitted to integrin-based
cell-matrix adhesions, so-called focal adhesions (FAs), where they
trigger biochemical signals feeding back to the actin
cytoskeleton. The main mechanism which has been suggested in this
context is the force-induced activation of the Rho signaling pathway
through guanine nucleotide exchange factors (GEFs) that reside in FAs
\cite{Geiger09}.  Activation of RhoA leads to activation of the
Rho-associated kinase (ROCK).  Active ROCK is able to phosphorylate
myosin light chain phosphatase (MLCP) to MLCP-P and thereby
deactivates the phosphatase. MLCP and MLCP-P are freely diffusible in
the cytoplasm and can reach the myosins in the SFs. Increased
phosphorylation of MLCP to MLCP-P by ROCK thus effectively leads to
increased phosphorylation of myosin light chain (MLC), increasing
myosin contractility along the SFs. In our model, this signaling
pathway is described by a system of reaction diffusion equations where
each enzymatic step is described by Michaelis-Menten kinectics
\cite{Besser07}. The actomyosin force is assumed to activate the
Rho-pathway with Michaelis-Menten kinetics. This is the simplest
assumption given a linear increase in biochemical activity at small
force and saturation at large force. The biochemical signal couples
into the SF contraction mechanics via the force velocity relation of
the myosin motors. In this way, a mechanical and biochemical positive
feedback loop is closed as depicted in \fig{fig_model}(b). The
mechano-chemical model has been introduced before to describe
inhomogeneous SF contraction upon myosin stimulation by the drug
calyculin \cite{Besser07}. To account for compliant substrates, we now
appropriately modify the boundary conditions of the SF equation. In
this situation, at each fiber end the traction forces $F_t$ exerted by
the SF have to be balanced by the elastic restoring forces of the
substrate. The latter is modeled as a linear elastic spring of
stiffness $k_s$. All model equations and parameter values are
summarized in the supplement.

\section{Bifurcation analysis}

Motivated by their experimental
relevance, we start with a bifurcation analysis with the
non-dimensional ratio of substrate and SF stiffness $k_s/k$ and the
stimulation strength $I$ as control parameters. The parameter $I$
accounts for the inhibitory effects of calyculin on MLCP: $I=1$
corresponds to an unperturbed system and $I>1$ ($I<1$) lead to higher
(lower) actomyosin activity. As state variable we introduce the
absolute value of the exerted traction forces, $F_t=|k_s u|$. An
alternative choice is the substrate deformation $u$, see supplemental
results.

\fig{fig_bif_f} shows the bifurcation diagram for the traction force
$F_t$ as a function of the stiffness ratio $k_s/k$ for different
values of the stimulation strength $I$. For each value of $I$, the
upper blue and the lower red branch represent the stable fixed points
for different values of the stiffness ratio $k_s/k$. The unstable
branch is indicated as dashed line. The S-shape of these curves
demonstrates that the system is strongly bistable due to the positive
feedback between force and signaling. The stable upper branch
corresponds to a highly contractile cell whereas the stable lower
branch corresponds to an inactive cell that fails to establish a
contractile state. Both bran{\-}ches increase monotonically with
substrate stiffness. However, the upper branch quickly saturates for
high stiffness ratios. It can be deduced from \fig{fig_bif_f} that for
all considered values of $I$ there exist a critical stiffness ratio,
defined by the left bifurcation point, below which the upper stable
branch vanishes. As a consequence, on very soft substrates, cells can
not establish a highly contractile state. Correspondingly on the stiff
side there exists a threshold above which unperturbed cells are always
contractile.

\begin{figure}
\includegraphics[width=0.7\textwidth]{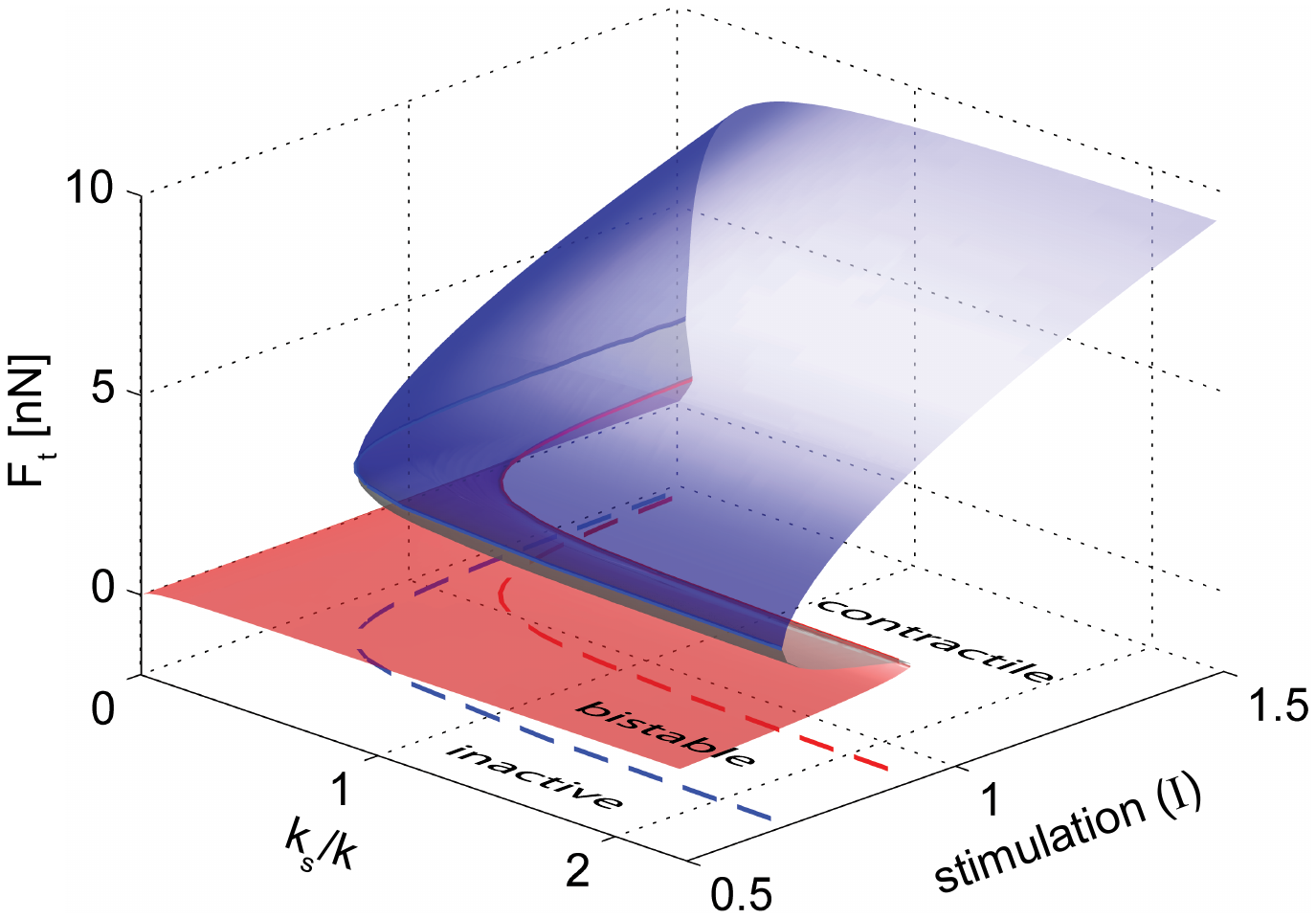}
\caption{Two parameter (stiffness ratio $k_s/k$, stimulation strength
  $I$) bifurcation diagram for the traction force $F_t$. Colored
  planes represent: stable upper branch (blue); stable lower branch
  (red); unstable branch (gray). A stability diagram is constructed by
  projecting the two bifurcation curves (red and blue solid lines)
  onto the parameter plane (dashed lines), subdividing it in
  contractile, bistable and inactive regions.}
\label{fig_cusp}
\end{figure}

By systematically sampling the 2D parameter space, the stable and
unstable branches become surfaces defined over the $(k_s/k)$--$(I)$
plane as shown in \fig{fig_cusp}. The two folds where the surface
bends over itself define two bifurcation curves representing the two
thresholds involved. Their projection on the parameter plane, shown as
dashed lines, yields the stability diagram of the system. The curves
divide the plane into three regions, namely inactive, contractile or
bistable. If the system is driven out of the bistable region by a
parameter change, it can be forced to a sudden transition between the
inactive and the contractile states. This prediction is consistent
with the experimentally observed abrupt appearance of mature SFs in
3T3 fibroblasts if the substrate stiffness is increased beyond a
threshold value of $\approx 3~\textrm{kPA}$ \cite{Yeung05}. Cells also
spread faster on stiff substrates \cite{Yeung05} which corresponds to
a faster buildup of forces as predicted by our model, see Fig.~S3.
The course of spreading experiments is indicated by upward arrows in
\fig{fig_bif_f}. However, they can not reach the upper stable branch
in the bistable region. Our results suggest that this state can be
attained by stimulated cells after washout of calyculin, as indicated
by the downward arrows in \fig{fig_bif_f}.

\section{Hysteresis and rate effects}

An important consequence of
the bistability revealed by our bifurcation an{\-}alysis is the
existence of a hysteresis loop for cell adhesion.  To construct
such a hysteresis cycle, cells had to be prepared on stiff substrates
in the highly contractile state, compare \fig{fig_bif_f}. By reducing
the substrate stiffness sufficiently slowly, such that the system can
adapt and remain in a quasi-steady state, it will follow the upper
stable branch until it reaches the left bifurcation point.  Here, the
upper branch becomes unstable and the system is forced into the
inactive state. When the control parameter is increased again, the
system will stay on the lower branch while it reaches the right
bifurcation point, where the lower branch becomes unstable. Thus, the
system is finally forced again onto the upper branch and the
hysteresis cycle is closed.

\begin{figure}[ht]
\includegraphics[width=0.7\textwidth]{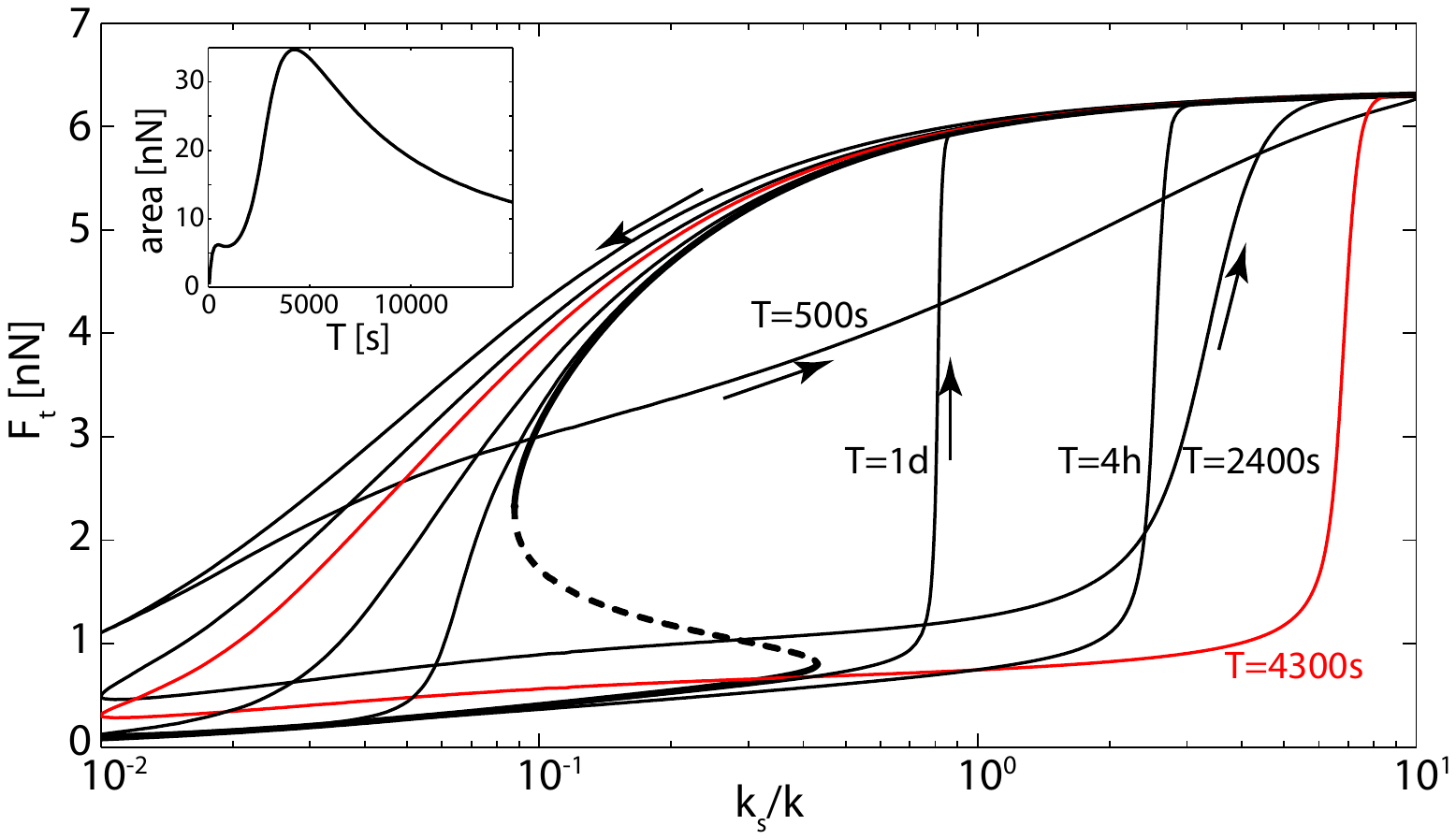}
\caption{Time course of traction force $F_t$ for cyclic varying substrate
stiffness with period $T\in\{500\,\textrm{s}, 2400\,\textrm{s}, 4300\,\textrm{s},
4\,\textrm{h}, 1\,\textrm{d}\}$, in the biochemically unperturbed case $I=1.0$.
The area of the hysteresis cycle as a function of the period (given as inset),
reaches a maximum at $T=4300\,\textrm{s}\pm 100\,\textrm{s}$ (red curve). The
bifurcation diagram (black, fat lines) is approached for $T\rightarrow\infty$.
Arrows indicate cycling direction.}
\label{fig_cyclic_f}
\end{figure}

The ideal hysteresis scenario is expected to occur only
for very slow chan{\-}ges in stiffness. In experiments, rate effects might
occur. To analyze this situation theoretically, we now can take full
advantage of our dynamical model. We have simulated this experiment
with COMSOL Multiphysics for a stiffness ratio $k_s(t)/k$ which
oscillates sinusoidally between $0.01$ and $10.0$ with different
periods $T$. This stiffness range is chosen such that it covers the
bistable region. The resulting time courses for the traction forces
are shown in \fig{fig_cyclic_f}. As an inset we show the encircled
hysteresis area as a function of the period $T$.  For very large
periods ($T=1\,\textrm{d}$), the system follows essentially the
bifurcation diagram. The area of the hysteresis cycle first increases
with the frequency and reaches a maximum for $T = 4300\,\textrm{s}$,
the red curve in \fig{fig_cyclic_f}. For very high frequencies the
hysteresis cycle tightens up again as the period of the mechanical
input becomes smaller than the relaxation time of the biochemical
system. In this case, the biochemical part of the system conserves the
initial activation and thus, the system can maintain higher forces on
soft substrates, leading to a rather flat time course of the traction
forces, compare $T=500\,\textrm{s}$ in \fig{fig_cyclic_f}.

\section{Conclusions and outlook}

A bifurcation analysis of our dynamical systems
model showed that the described positive mechano-chemical feedback
cycle leads to bistability for contraction as a function of substrate
stiffness. This readily explains the experimentally observed abrupt
change in the morphology and contractility of fibroblast-like cells at
a rigidity threshold. Due to its dynamical nature,
our model also allows us to predict the different hysteresis curves
expected for different frequencies of dynamically changing substrate
stiffness. Recently different experimental setups have been
introduced which make cell experiments on substrates with
time-dependent stiffness possible. Hydrogels made
from thiolated hyaluronic acid and polyethylene glycol diacrylate
exhibits a time-dependent increase in the Young modulus
\cite{Engler07}. The stiffness can be reduced again by breaking formed
disulfide bonds with dithiothreitol. Alternatively, micromanipulation
systems like AFMs or microplates can be used to mimic compliant substrates
through an electronic feedback system \cite{Mit09}.

Our analysis demonstrates that such dynamical protocols open up
a new dimension of controling cell behaviour through physical cues.
Here, we have restricted our treatment to a single fiber as a paradigm
as to how cells couple biochemistry and mechanics. Future work should
also address how different fibers interact with each other. Fibers in the same cell
should share the biochemical input through diffusion fields, while all fibers
(including those from other cells) might interact mechanically through a compliant
substrate.

\acknowledgments{We thank F. Rehfeldt for helpful discussions. This work was supported 
by the cluster of excellence CellNetworks at the University of Heidelberg.}

\begin{figure}
\includegraphics[width=0.7\textwidth]{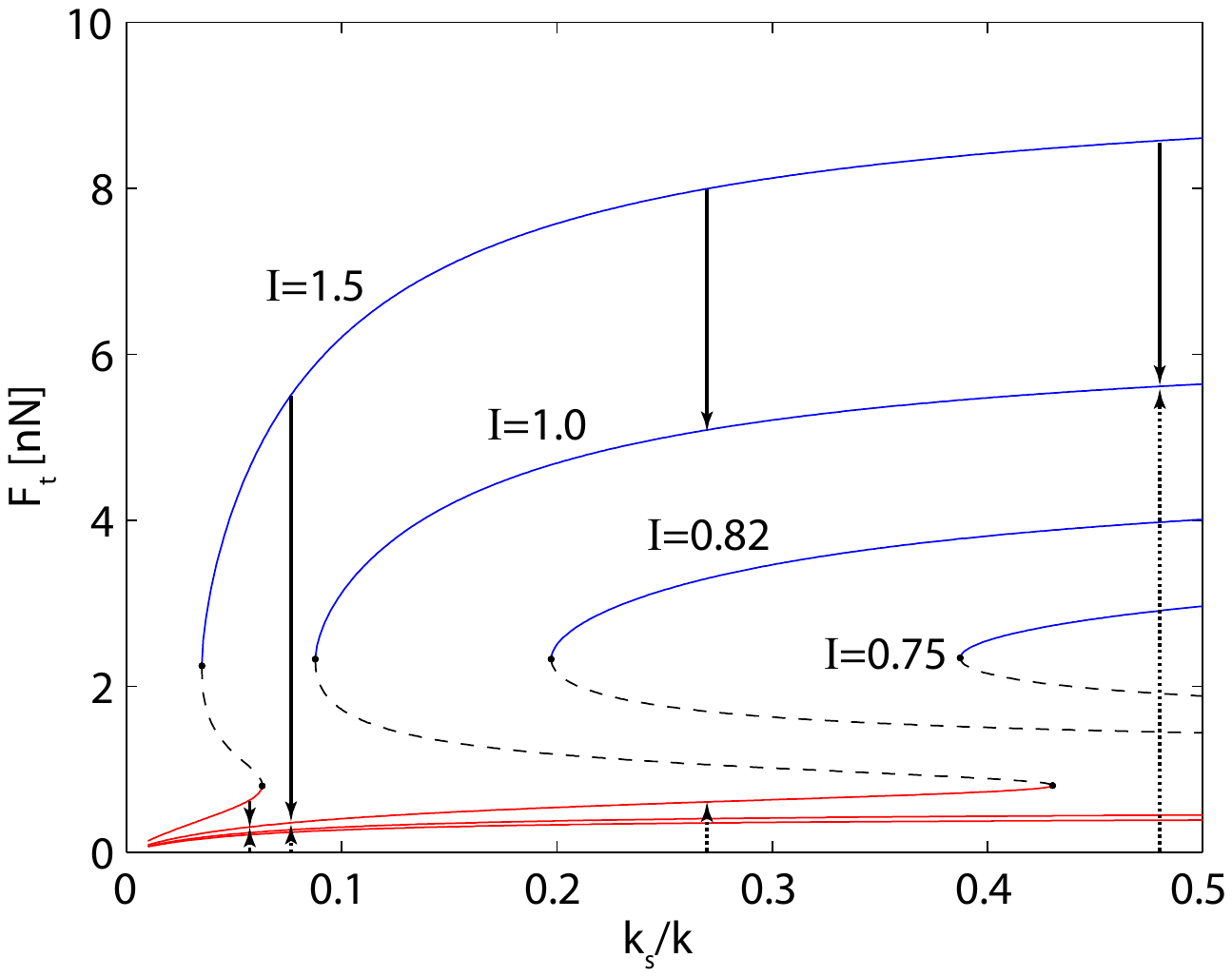}
\caption{Bifurcation diagrams for the traction force $F_t$ using the
  stiffness ratio $k_s/k$ as control parameter and varying stimulation
  strength $I\in\{0.75, 0.82, 1.0, 1.5\}$. Shown are: stable upper
  branches (blue) that correspond to contractile states; stable lower
  branches (red) that correspond to inactive states; unstable branches
  (dashed lines); saddle node bifurcation points (black dots). Upward
  and downward arrows illustrate spreading and calyculin washout
  experiments, respectively.}
\label{fig_bif_f}
\end{figure}

\end{document}